\documentstyle[prd,aps,epsf]{revtex}      
\begin{document}

\title{Parametric amplification of waves during gravitational
collapse: a first investigation}

\author{Stanislav V. Babak $^{1}$ and Kostas Glampedakis$^{1,2}$ }

\address{$^1$ Department of Physics and Astronomy,
Cardiff University,  Cardiff CF2 3YB, United Kingdom}
\address{$^2$ Department of Mathematics, 
University of Southampton, Southampton SO17 1BJ, United Kingdom\\
{\rm E-mail: babak@astro.cf.ac.uk ;  glampedakis@astro.cf.ac.uk}}

\maketitle

\begin{abstract}

We study the dynamical evolution of perturbations in the  
gravitational field of a collapsing fluid star. Specifically, we consider 
the initial value problem for a massless scalar field in a spacetime similar 
to the Oppenheimer-Snyder collapse model, and numerically evolve 
in time the relevant wave equation. Our main objective is to examine whether 
the phenomenon of parametric amplification, known to be responsible for 
the strong amplification of primordial perturbations in the expanding 
Universe, can efficiently operate during gravitational collapse. Although 
the time-varying gravitational field inside the star can, in principle, 
support such a process, we nevertheless find that the perturbing field 
escapes from the star too early for amplification to become significant. 
To put an upper limit in the efficiency of the amplification mechanism 
(for a scalar field) we furthermore consider the case of perturbations 
trapped inside the star for the entire duration of the collapse. 
In this extreme case, the field energy is typically amplified at the level 
$\sim 1\%$ when the star is about to cross its Schwarszchild radius. 
Significant amplification is observed at later stages when the star has 
even smaller radius. Therefore, the conclusion emerging from our simple 
model is that parametric amplification is unlikely to be of significance during gravitational collapse. Further work, based on more realistic collapse models, 
is required in order to fully assess the astrophysical importance of parametric amplification.

\end{abstract}


\section{Introduction}

\subsection{A brief background}

The gravitational collapse of stellar objects with mass over a certain 
threshold is known to produce one of the most spectacular events in Universe. 
Due to the release of an enormous amount of energy, collapse is one of most 
favourable sources for all kinds of electromagnetic observations. It is 
expected (or hoped) that the same will come true for gravitational wave observations, which are about to begin in the next few years with the completion of a 
network of interferometric detectors (for a 
recent review see \cite{review}). A considerable amount of theoretical 
work (numerical in its majority) has been devoted to the description/prediction of 
the emitted gravitational waveform produced by such a 
catastrophic event (see for instance \cite{Muller}). Despite this effort, 
great uncertainties still remain, mainly due to the unknown initial state 
of the collapsing body, the complexity of the physics involved in realistic collapse 
and the related excessive computational requirements. 

The simplest general relativistic collapse model was introduced more that sixty 
years ago with the seminal work of Oppenheimer and Snyder (hereafter O-S) 
\cite{op}. In that model the collapsing star is approximated as a freely 
falling ``dust'' fluid ball (see Section II for more details). 
Using the O-S model, Price \cite{price} was the first to study the 
evolution of scalar field perturbations during spherically symmetric 
gravitational collapse, focusing on the last stages where a black hole is 
formed. Subsequent studies by Cunningham, Price and Moncrief \cite{cunningham} were extended to gravitational perturbations. It was found that, at 
the final stage of the collapse, the emitted waveform was dominated by the 
exponentially decaying quasinormal mode ``ringing''  of the newly born 
black hole. At even later times, the signal decayed as a power-law, 
representing radiation backscattered from the asymptotic gravitational
field. 

Collapse of relativistic polytropes was modelled by Seidel \cite{seidel} 
(for a non-rotating configuration) and by Stark and Piran \cite{stark} 
(for an axisymmetric rotating configuration), who provided an estimate of the 
amount of energy released via gravitational radiation. In their calculations,
the resulting waveform was mainly comprised of a short burst followed by the
black hole ringdown. Recent efforts include the relativistic 
rotational core-collapse study of Dimmelmeier et.al. \cite{harald} and 
the more general investigation by Fryer et.al. \cite{fryer}. We also 
refer the reader to the article of Font \cite{font} (and references therein) 
where numerical techniques in general relativistic hydrodynamics are
discussed.


\subsection{Motivation}        

In the present paper we aim to study a specific aspect of gravitational
collapse that (to the best of our knowledge) has been overlooked in the
literature, namely, the possibility of having parametrically amplified 
perturbations within the star. That such a process could operate, at least 
in principle, can be justified with the following argument.

It is well known that there is a qualitive similarity between the 
time-dependent gravitational field of a spherically symmetric collapsing 
star and the gravitational field of the closed Friedmann cosmological model. 
This similarity has been quantified in the
celebrated O-S collapse model \cite{op}, which, despite its simplicity,
gives reliable results. On the other hand, one of the most important results 
in the study of cosmological perturbations is that their interaction with the 
background gravitational field of the expanding Universe can lead to considerable 
amplification \cite{lpg} (in addition to the overall ``adiabatic'' 
change $\sim a^{-1} $,  where $a$ is the cosmological scale factor). 
This mechanism is believed to be responsible for the immense amplification of 
tiny primordial gravitational quantum vacuum fluctuations and their promotion 
to classical fields, which are likely to have left measurable imprints 
in the CMB radiation observed today \cite{cmbr}. Moreover, these fields 
put a strong candidature for direct observation by the gravitational wave detectors \cite{review}.

Hence, based on the two points above, a natural question that arises is 
whether the gravitational field of a collapsing star could pump energy 
into perturbations (that can be of fluid or gravitational nature) temporary 
``living'' inside the star.
In this work we plan to give an answer to this question by considering the simplest
possible model for a collapsing star. In particular, we keep those pieces of physics
which are related to the phenomenon we try to study, and strip off unrelated features
which might be a cause of technical problems and confusion. We shall adopt a collapse
model which very closely resembles the known O-S model, but is even more simplified
regarding the exterior to the star gravitational field (see next Section).
In a realistic situation, we can think of a collapsing star as a generator of all 
kinds of perturbations (fluid, gravitational). In our toy model, this process is 
``mimiced'' by placing some initial field inside the star and studyingits subsequent
evolution. Moreover, we are free to choose the time at which this field
is first placed inside the star. Finally, another major simplification is to consider 
a massless scalar field as the perturbation. We expect, however, that our results 
can be extrapolated to the realistic case of gravitational perturbations. This 
claim is based on the work of Ford and Parker \cite{ford}, where both 
scalar and gravitational perturbations in Friedmann spacetime were considered.

The remainder of the paper is organised as follows. In Section II we give a
brief review of the O-S collapse model (subsection IIA) and subsequently discuss 
the properties of the scalar wave equation in the field of a collapsing star and
of full Friedmann spacetime (subsection IIB). In subsection IIC we introduce our
simplified collapse model. Section III is devoted to our numerical results 
(time-evolutions). Analytical calculations regarding the amount of amplification 
in terms of the field's energy, are presented in Section IV. Section V offers a 
physical insight into the results of the previous Sections. Finally, a concluding 
discussion can be found in Section VI. Two Appendices are devoted to some technical 
details. Throughout the paper we have adopted geometrised units $c=G=1$.


\section{A toy-model for gravitational collapse}

\subsection{The Oppenheimer-Snyder model}

We set off by giving a brief review of the standard O-S model (for a detailed
presentation see \cite{MTW}). In this model, the spacetime of a collapsing, 
spherically symmetric, pressureless, homogeneous fluid ``ball'' is constructed
by patching together a piece of Schwarzschild geometry (describing the 
vacuum outside the star) and a piece of closed ($k=+1$) Friedmann geometry 
(describing the stellar interior). The two metrics are matched smoothly across 
the surface of the star. Explicitly, the interior metric is written in familiar 
comoving coordinates $(\eta,\chi,\theta,\phi)$,
\begin{equation}
ds^2= a^2(\eta) [ -d\eta^2 + d\chi^2 + \sin^{2}\chi ( d\theta^2 +
\sin^{2}\theta d\phi^2) ]
\label{fmetric}
\end{equation}
while the exterior metric is, in Schwarzschild coordinates,
\begin{equation}
ds^2= -f(r) dt^2 + f^{-1}(r) dr^2 + r^2 (d\theta^2 +
\sin^{2}\theta d\phi^2)~,
\label{smetric}
\end{equation}
where $ f(r)= 1- 2M/r$. The scale factor in the Friedmann domain is,
\begin{equation}
a(\eta)= \frac{1}{2} a_{\rm m} ( 1 + \cos\eta )
\label{scale}
\end{equation}
with $ a_{\rm m } $ denoting its initial value (which is also the maximum 
value). The surface of the star is labelled as $\chi_{\rm o}$ and $R$ in these two
coordinate frames and as seen in Schwarzschild coordinates is moving inwards as 
a timelike radial geodesic with zero initial velocity at some distance $R_{\rm i}$
corresponding to the initial stellar radius at the onset of collapse. This translates into
\begin{equation}
R(\eta)= \frac{1}{2} R_{\rm i} ( 1 + \cos\eta )~.
\label{Rad}
\end{equation}
The relation between the two different time coordinates is given by 
\begin{equation}
t= 2M \ln  \left | \frac{ (R_{\rm i}/2M -1)^{1/2} + \tan(\eta/2)}
{(R_{\rm i}/2M -1)^{1/2} - \tan(\eta/2)} \right | +
2M(R_{\rm i} -1)^{1/2} [ \eta +(R_{\rm i}/4M)(\eta + \sin\eta )]~.
\label{teta}
\end{equation}
Note that the conformal time coordinate $\eta $ is restricted between
$\eta=0$ (initial time when collapse begins) and $\eta=\pi$ at which moment
the star has shrank to zero radius. Of course, the physically relevant
value $\eta_c$ for $\eta$ is the time when the star is crossing 
its Schwarzschild radius,
\begin{equation}
\eta_c=  \arccos \left ( \frac{4M}{R_{\rm i}} - 1 \right )~.
\label{eta_c}
\end{equation}  
The initial stellar radius as measured in units of the stellar total mass $M$ is,
\begin{equation}
\frac{R_{\rm i}}{M}= 2\sin^{-2}\chi_{\rm o}~.
\label{RM}
\end{equation}
The parameter $\chi_{\rm o}$ can take any value between $0$ 
($R_{\rm i}/M \to \infty$) and $\pi/2$ ($R_{\rm i}/M = 2$). 
A typical collapsing object has $R_{\rm i} \sim 10^4 -10^5 M $
(the size of a white dwarf), and this value can rise even higher for
supermassive stars.


\subsection{The scalar wave equation in the field of a collapsing
             star}

As we mentioned earlier in this paper, we shall consider the simplest
possible case of perturbations, i.e. a massless scalar field $\Phi$, 
which obeys the covariant wave equation $ \nabla^{\alpha}\nabla_{\alpha}
\Phi =0 $. We can 
immediately separate the angular dependence by means of the decomposition
\begin{equation}
\Phi= \sum_{\ell,m}R_{\ell m} Y_{\ell m}(\theta,\phi)~.
\label{decomp}
\end{equation}
In the Friedmann domain (also simply termed ``interior'' hereafter), the 
wavefunction $R_{\ell m}$ can be rescaled as
\begin{equation}
R_{\ell m}(\eta,\chi)= \frac{\psi_{\ell m}(\eta,\chi)}{a(\eta) \sin\chi}~.
\end{equation}
Note that by this operation we have isolated the usual ``adiabatic''
change $\sim a^{-1}$ due to the contracting background.
The new wavefunction $\psi$ satisfies (for brevity we drop the $\ell,m$
subscript hereafter),
\begin{equation}
\partial_{\eta}^2 \psi -\partial_{\chi}^2 \psi + V(\eta,\chi) \psi =0~,
\label{master1}
\end{equation}
where the effective potential $V$ is given by (for $\chi \leq \chi_{\rm o} $),
\begin{equation}
V(\eta,\chi)= \frac{\ell(\ell +1)}{\sin^2\chi} -1 -\frac{\ddot{a}}{a}
           = \frac{\ell(\ell +1)}{\sin^2\chi} -\frac{1}{2} \cos^{-2}(\eta/2)~.
\label{pot1}
\end{equation}   
Similarly, for the Schwarzschild domain (to be called ``exterior'') 
we have,
\begin{equation}
R(t,r)= \frac{1}{r} \psi(t,r)
\end{equation}
with $\psi$ satisfying the familiar Regge-Wheeler equation,
\begin{equation}
\partial_{t}^2 \psi - \partial_{r_{\ast}}^2 \psi - V(r) \psi =0 
\label{RW}
\end{equation}
with
\begin{equation}
V(r)= f(r) \left ( \frac{\ell(\ell +1)}{r^2} + \frac{2M}{r^3} \right )~,
\label{pot2}
\end{equation}
and where $r_{\ast}$ is the so-called ``tortoise'' radial coordinate
defined by $dr_{\ast}/dr= 1/f(r)$.  
 
Naturally, our attention will be mainly focused on the interior 
time-dependent field which is identical (by construction) to the
field of a closed Friedmann model. The exterior field is of
little concern to us, as it is responsible for effects like the
quasinormal mode ringing and late time power-law  tails which are the
expected (and well studied) dominant components of the emitted signal 
at the final stages of a realistic collapse scenario.

\begin{figure}[tbh]
\centerline{\epsfysize=8cm \epsfbox{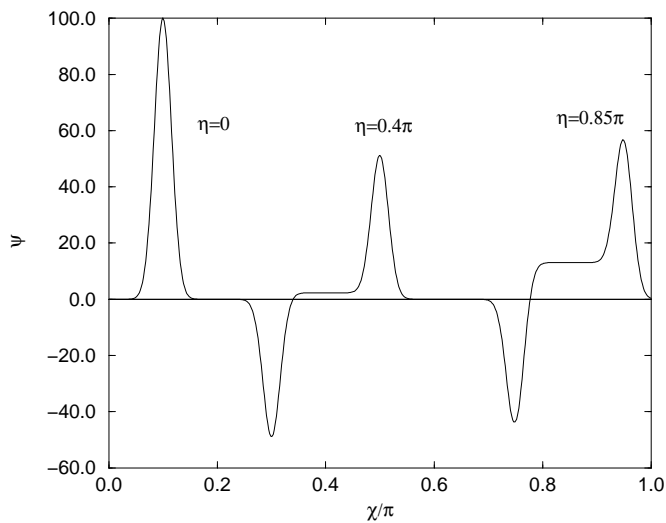}}
\caption{Scalar field propagation in Friedmann
spacetime. The initial field is shown on the left and later
``snapshots'' are shown on the center of the figure ($\eta=0.4\pi$) and
on the right ($\eta=0.85\pi$).} 
\label{fig1}
\end{figure}

As a warm up, we first consider the initial value problem for the
wave equation (\ref{master1}) in a pure Friedmann gravitational field.
Moreover, we restrict ourselves to the monopole case $\ell=0$. 
Assuming  static initial data $\psi(0,\chi)= \psi_{\rm o}(\chi)$,
$\partial_{\eta} \psi (0,\chi)= 0$ and imposing the additional 
boundary condition $\psi(\eta,\pi)=0$, the general solution of 
(\ref{master1}) is given by,
\begin{equation}
\psi(\eta,\chi)=  \frac{2}{\pi}\sum_{n=1}^{+\infty} 
 \sin(n\chi) \int_{0}^{\pi}d\chi~ \psi_{\rm o}(\chi) 
\sin(n\chi)
\left [ \cos(n\eta) + \frac{1}{2n} \tan(\eta/2) 
\sin(n\eta) \right ]~.
\label{solution1}
\end{equation}
The key feature of eqn. (\ref{solution1}) is the term $\sim \tan(\eta/2)$.
This term originates from the $\ddot{a}/a$ term in the potential 
eqn. (\ref{pot1}) and consequently encodes all the effects associated with the time 
variance of the background gravitational field. 
We can apply the solution (\ref{solution1}) for initial data of the  form of a 
narrow Gaussian pulse. The resulting field is shown in Fig.~\ref{fig1}.

The field initially evolves as in flat spacetime, but at later times,
we clearly observe a growth in the amplitude which is more pronounced
in the ``wake'' behind the travelling pulse. This wake is mainly comprised
of low-frequency waves (which ``feel'' more strongly the background 
gravitational field) as opposed to the main pulse which includes the
higher frequency contribution. As $ \eta $ approaches $\pi$ the 
field becomes enormous as compared to its initial amplitude and eventually diverges,
as predicted by eqn. (\ref{solution1}). This behaviour is a typical example of 
parametric amplification in Friedmann spacetime.


\subsection{Pushing the simplification further}

\begin{figure}[tbh]
\centerline{\epsfysize=5.5cm \epsfbox{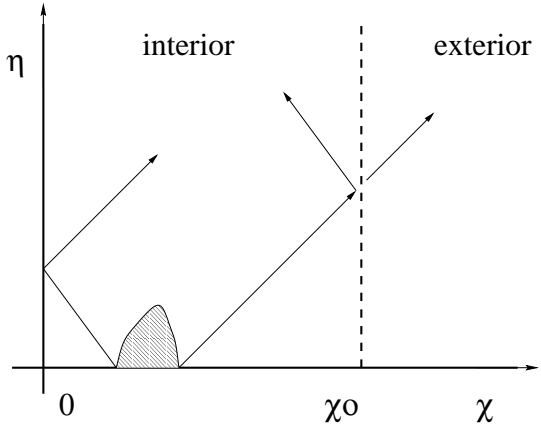}}
\vspace{0.3cm}
\caption{A schematic illustration of our problem: Initial data
placed inside a ``collapsing'' star propagate inwards and outwards
experiencing partial and total reflections at the stellar surface 
($\chi=\chi_{\rm o}$) and at the center ($\chi=0$) respectively.}  
\label{scheme}
\end{figure}

Let us now return to the O-S model. The difference in the field evolution 
from the preceding picture is obvious: the field, as it propagates outwards, 
will reach the stellar surface at $\chi_{\rm o}$. Once there, it will 
partially transmit to the exterior and partially reflect back to the interior
(see Fig.~\ref{scheme}). In principle, one could try and describe such a process analytically, 
by making some simplifications for the exterior field (for example, by 
keeping only the centrifugal term in the potential). However, such an 
attempt soon runs into serious technical problems related to the coordinate ``discontinuity'' at the stellar surface. To overcome this difficulty, 
one could use comoving Novikov coordinates \cite{MTW} which smoothly covers 
the entire spacetime. However, 
in this case the resulting exterior wave equation is not separable, 
an undesired feature for any analytical work. At the same time, we should 
be almost certain that this coordinate-induced 
problems has nothing to do with the physics we try to study here.
Of particular relevance to this point is the recent study by
Laarakkers and Poisson \cite{poisson} on scalar wave propagation in the 
so-called Einstein-Strauss spacetime. This spacetime is, as the O-S model, 
a combination of Schwarzschild and Friedmann geometries but patched in the 
opposite sense with the interior being Schwarzschild and the exterior Friedmann.   
Apart from providing fully numerical results, these authors considered 
a simple toy-equation that closely resembles the pair of exact wave equations 
(\ref{master1}), (\ref{RW}). The construction was based on simply replacing 
eqns. (\ref{master1}),(\ref{RW}) with a single wave equation by extending the 
Friedmann coordinates over the entire spacetime. It was shown that the simplified
wave equation captures most of the effects present in wave propagation in the
Einstein-Strauss spacetime.

For the purposes of the present paper we have chosen to follow a similar 
path and replace eqns. (\ref{master1}),(\ref{RW}) with the toy-equation,
\begin{equation}
\partial_{\eta}^2 \psi - \partial_{\chi}^2 \psi + V(\eta,\chi) \psi =0~,
\label{master2}
\end{equation}
where the new effective potential is,
\begin{equation}
V(\eta,\chi) = 
\cases{ 0 
\,& for $\chi > \chi_{\rm o},$ ~~ \mbox{exterior} 
\cr
\cr
-\frac{1}{2}\cos^{-2}(\eta/2)  \,&
for $ \chi \leq \chi_{\rm o},$~~~ \mbox{interior}  \cr}
\label{potnew} 
\end{equation}
Hereafter, we shall consider only the monopole case $\ell=0$. We have verified
that all results presented in this paper remain qualitively the same even for 
higher multipoles. The important point is that the new potential 
is {\em identical} to the ``exact'' O-S potential in the interior, where 
the phenomenon of parametric amplification would take place. On the other
hand, our model is unable to describe any effects related to the exterior
gravitational field, such as backscattering of outgoing waves. In principle, 
this last feature could be of importance, as it would lead to trapping
of some portion of the initial field inside the star. For this reason,
we have also considered a star whose surface fully reflects any outgoing waves
(see Section IV).    

In a sense, eqn. (\ref{master2}) describes a 
static star (as seen by both external and internal observers) with some kind 
of time-varying gravitational potential in its interior. This star, and with 
respect to external observers {\em only}, knows nothing about collapse. 
Physical parameters like the total duration of collapse have to be ``borrowed'' 
from the O-S model. We believe that the wave equation (\ref{master2}), just as 
in the case of Einstein-Strauss spacetime, yields reliable predictions
on scalar wave dynamics.

It is possible to find an analytic solution to the above toy-equation (see \cite{poisson} for a treatment of the $k=0$ Friedmann equivalent) that 
describes the first transmission/reflection of a wavepacket through the potential discontinuity. In principle, we could extend this solution to incorporate multiple transmissions/reflections for the needs of our problem. However, the resulting expression becomes increasingly messy. This is the main reason that led us to 
approach the problem numerically.


\section{Time evolutions}

\begin{figure}[tbh]
\centerline{\epsfysize=8cm \epsfbox{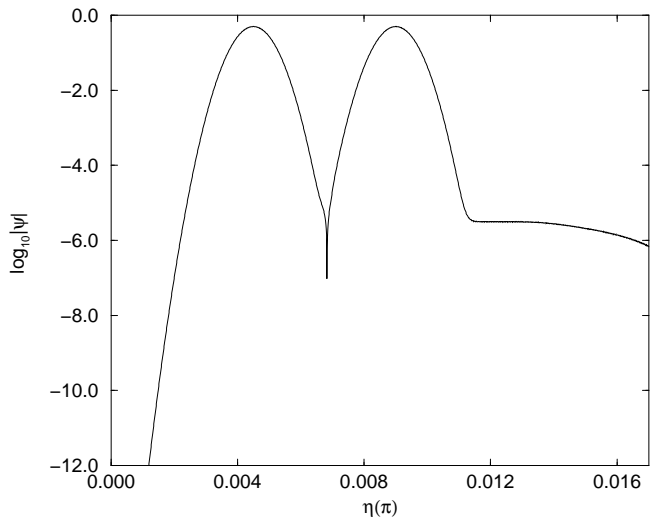}}
\vspace{0.2cm}
\caption{The field as observed at the radial location $\chi= 1.5 \chi_{\rm o}$,
for $R_{\rm i}= 10^4 M$. The initial pulse was placed at $\chi_{\rm o}/2$ and 
$\eta=0$.} 
\label{t_ev0}
\end{figure}

\begin{figure}[tbh]
\centerline{\epsfysize=8cm \epsfbox{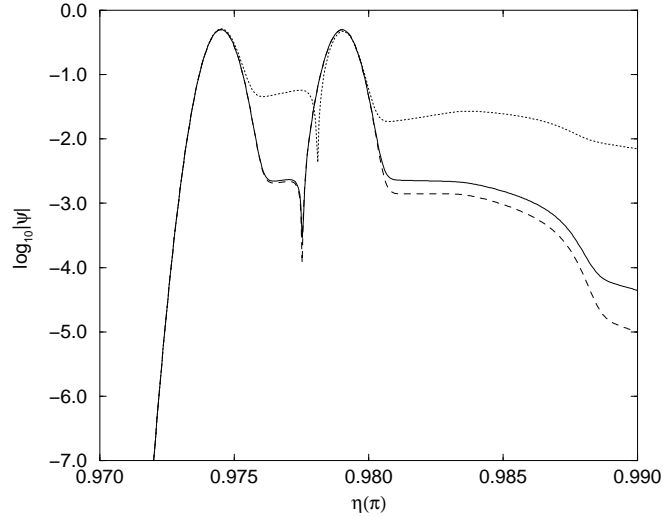}}
\vspace{0.2cm}
\caption{The field (solid curve) as observed at $\chi= 1.5 \chi_{\rm o}$, for
$R_{\rm i}= 10^4 M$. The initial pulse was placed at $\eta_i= 0.97\pi $
(corresponding to a radius $R \approx 22M$). The dashed and dotted curves 
depict the field for a ``frozen'' static potential $V(\eta=\eta_i)$ and 
$V(\eta=\eta_f)$ respectively.} 
\label{t_ev1}
\end{figure}

\begin{figure}[tbh]
\centerline{\epsfysize=8cm \epsfbox{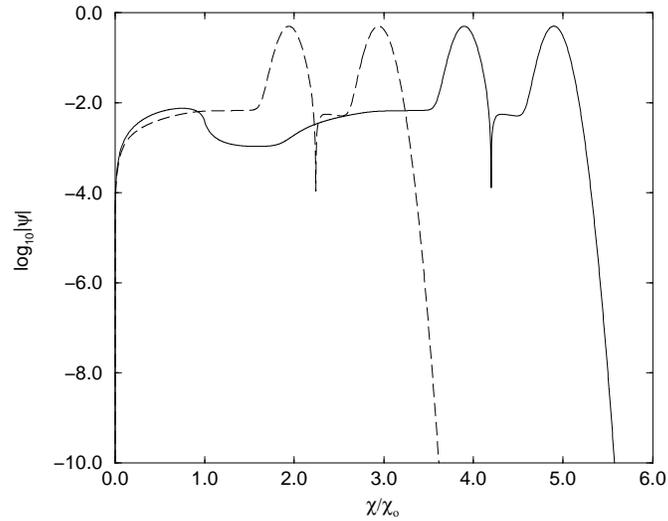}}
\caption{``Snapshots'' of the field at time $\eta= 0.9998\pi$ (solid curve) 
corresponding to a radius $R \approx 0.001M$ and at $\eta= 0.99\pi$ 
(dashed curve) corresponding to $R \approx 2M $. In the former, it is easy to 
distinguish the amplified field in the stellar interior. The initial radius
is $R_i= 10^4 M$, and the initial data were placed at $\eta_i =0.98\pi$.} 
\label{t_ev3}
\end{figure}

\begin{figure}[tbh]
\centerline{\epsfysize=8cm \epsfbox{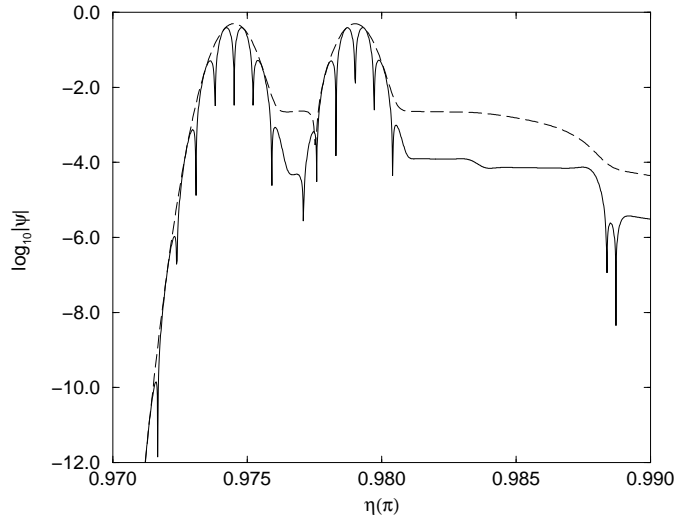}}
\caption{Dependence of the observed signal (at $\chi= 1.5\chi_{\rm o}$) on 
the frequency content of the initial data: the dashed (solid) curve was generated by  a ``simple'' (modulated) Gaussian pulse as initial data. 
It is obvious that high frequency components escape from the star 
much easier.} 
\label{t_ev2}
\end{figure}

We have written a simple numerical code for solving eqn. (\ref{master2}) after
prescribing some initial data (typically in the form of a Gaussian pulse).
The code, which employs a standard leapfrog step algorithm, was tested for
convergence and stability. We have widely experimented with collapsing models
of different $R_{\rm i}$ and placing the initial data at different stages of the
collapse. A typical example of our results for the field 
as observed outside the star is shown in Fig.~\ref{t_ev0}. In this simulation we 
placed the initial data at $\eta=0$, i.e. at the onset of the collapse, and 
centered at $\chi= \chi_{\rm o}/2$, while the initial radius is $R_{\rm i}= 10^4 M$. 
We see that the field has escaped from the star, essentially unaffected, after
a lapse of time of the order $ \sim 3\chi_{\rm o}/2 $ (which is the light 
time-travel for the, initially ingoing, wave component to reach the surface). 
Only a minute fraction of the initial field (specifically the low frequency 
components) has been left inside the star while $R \gg M $.  
The same scenario is repeated in Fig.~\ref{t_ev1}, but now 
the initial data is placed at a later time $\eta_i= 0.97\pi$.
This corresponds to an instantaneous stellar radius $R \approx 22M $. The 
final time is $\eta_f \approx 0.99\pi$ which gives a radius 
$R \approx 2.1M$. In this case, the potential inside the star is much more 
pronounced and consequently is more effective in trapping inside the star
a small fraction of the initial pulse for a considerable time. 
The majority of the field, however, escapes during the first two crossings of the stellar surface. In the same Figure we illustrate the signal from a star with a potential ``frozen'' at its value at $\eta= \eta_i$ (dashed curve) and 
at $\eta= \eta_f $. We can see that the general appearance of the signal 
is similar for a static and
non-static potential. It is very hard to say if there are any genuine features 
of parametric amplification present in the signal. For this reason we illustrate (see 
Fig.~\ref{t_ev3}) the field as it looks at $\eta= 0.998\pi$ when the stellar 
radius has decreased to $R \approx 0.001M$. An amplified field inside the star
can be clearly seen now (although it could never escape to infinity).
In all the situations we examined, we reached a similar conclusion. Amplification
becomes notable only after the star has crossed the Schwarzschild radius.
Our final time evolution concerns the appearance of the emitted field as a 
function of the frequency content of the initial data. This can be simply
done by considering as initial data a modulated Gaussian pulse,
\begin{equation}
\psi_{\rm o}(\chi)= \exp[ -0.5(\chi-\chi_{\rm c})^{2}/\sigma^2] \sin[\lambda
(\chi-\chi_{\rm c})]~,
\end{equation}
where $\chi_{\rm c} $ the pulse location and $\lambda $ is the modulation
frequency.  In Fig.~\ref{t_ev2} we show a comparison between the
signals originated by an unmodulated pulse and a pulse with 
$\lambda= 20/\chi_{\rm o}$. All other parameters are the same to the ones
of Fig.~\ref{t_ev1}. As we should have expected, the high-frequency
pulse is escaping from the star more easily, and the late time ``tail''  is
strongly suppressed.  

\pagebreak

\pagebreak

\section{A ``mirror'' star}

The results of the previous Section are clearly not very optimistic
regarding the efficiency of parametric amplification. In order to put an
upper limit to this efficiency (at least for scalar perturbations) we have 
considered a ``mirror'' star, i.e. a star which totally reflects any outgoing 
field at the surface. Clearly, this situation is the most favoured one from the point 
of view of the amplification process, as the field will stay trapped inside the star 
for the entire collapse event. 

Mathematically, this mirror star is modelled by setting the additional 
boundary condition $\psi(\eta,\chi_{\rm o})= 0$. Hence, the field inside the star 
has two inflection points at $\chi= 0, \chi_{\rm o}$. It is straightforward 
to find an analytic solution to this problem that also satisfies static initial 
data $\psi(0,\chi)= \psi_{\rm o}(\chi)$ (see Appendix A for  details),
\begin{equation}
\psi(\eta,\chi)= \frac{2}{\chi_{\rm o}}
\sum_{n=1}^{+\infty} C_{\rm n} \sin(n\pi \chi/\chi_{\rm o}) 
\left [ \cos(n\pi \eta/\chi_{\rm o}) + \frac{\chi_{\rm o}}{2n\pi} \tan(\eta/2) 
\sin(n\pi\eta/\chi_{\rm o}) \right ]~,
\label{solution2}
\end{equation}
where
\begin{equation}
C_{\rm n}= \int_{0}^{\chi_{\rm o}}d\chi \psi_{\rm o}(\chi) 
\sin(n\pi\chi/\chi_{\rm o})~.
\end{equation}
We can strictly quantify the amplification of the field by means of the total
energy
\begin{equation}
E= \frac{1}{2}\int_{0}^{\chi_{\rm o}} d\chi 
\left [ (\partial_{\eta} \psi)^2 + (\partial_{\chi} \psi)^2 + V \psi^2
\right ]~.
\label{energ1}
\end{equation}
It can be easily shown (see Appendix B) that 
\begin{equation}
E(\eta) -E(0)= \frac{1}{2} \int_{0}^{\eta} \int_{0}^{\chi_{\rm o}} 
d\eta^{\prime} d\chi~ \psi^2 \partial_{\eta^{\prime}} V~.
\label{energ2}
\end{equation} 
In this simple way we can see that the field energy is not conserved due
to the time-varying potential.

\begin{figure}[tbh]
\centerline{\epsfysize=8cm \epsfbox{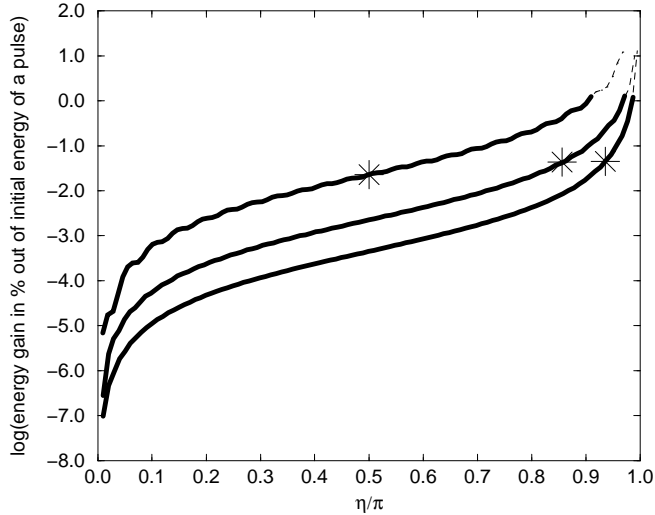}}
\caption{Setting an upper limit for the amplification factor: 
We plot the percentage fractional amplification as a function of time. 
Three different initial radii are shown: $R_{\rm i}= 100M$ (top curve),
$R_{\rm i}= 1000M$ (middle curve) and $R_{\rm i}= 5000M$ (bottom curve).
Different stages of the collapse can be distinguished in the following way:
The solid curves terminate at the points that correspond to horizon crossing
by the stellar surface ($R=2M$). The dashed curves continue down to 
$R \approx 0.25 M$. The stars denote the times where the star has a 
radius $R= 50M$.} 
\label{figmir1}
\end{figure}

Combining eqn.~(\ref{solution2}) with eqn.~(\ref{energ2}), we can monitor the energy
of the scalar field inside the collapsing star. 
Results of this calculation, for a selection of $R_{\rm i}$, are shown in 
Fig.~\ref{figmir1}, where we plot the quantity $\log_{10} |100 \times \{E(\eta) -E(0)\}/E(0)|$
(i.e. the percentage fractional change in the energy) as a function of
$\eta$. The conclusion from this calculation is rather dissapointing
from an ``observational'' point of view: The amplification factor is $\sim 1\%$ 
when the star is about to cross its Schwarzschild radius. 
It only becomes important, at later times, when the radius is close to zero.
This is in agreement with the behaviour seen in the time-evolutions in the field
of a ``transparent'' star (Section III). According to 
Fig.~\ref{figmir1}, there is a $\sim 10 \%$ amplification when $R \sim 0.25M$.
It is interesting to note that, according to Fig.~\ref{figmir1}, the amplification
factor is almost the same for a given value of $R$ and different $R_{\rm i}$. 
An intuitive explanation of this behaviour is given in the following Section.
It is important to mention that the amplification factor at the moment of horizon 
crossing proved to be sensitive to the width of the initial pulse. Specifically,
by decreasing the width of the pulse the amplification factor can drop down by 
an order of magnitude. This can be understood in terms of the fact that parametric amplification strongly depends on the frequency (see discussion in the next Section).

\begin{figure}[tbh]
\centerline{\epsfysize=8cm \epsfbox{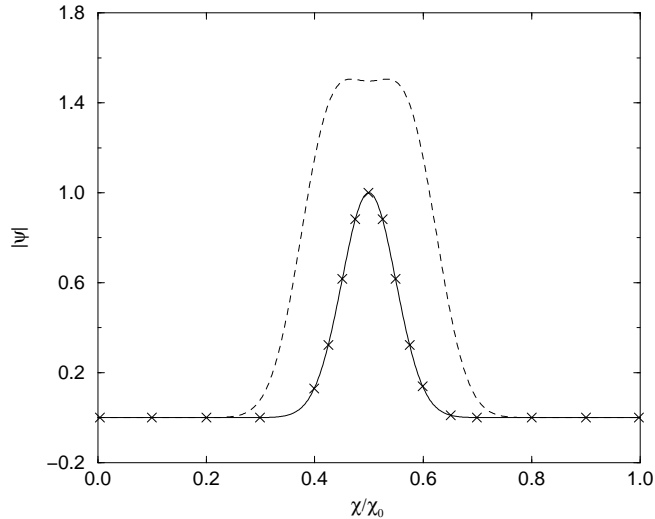}}
\caption{Different ``snapshots'' of scalar field propagating inside
a ``mirror'' star. The initial radius of the star is $R_{\rm i}= 100M$. 
The initial pulse is plotted with a solid curve, the field at $\eta=0.7\pi$
(at which moment has reached the maximum value) is denoted by crosses, 
and finally, the strongly amplified field at $\eta= 0.998\pi$ is denoted by 
the dashed curve.} 
\label{figmir2}
\end{figure}

In Fig.~\ref{figmir2} we show snapshots of  the field itself inside the 
``mirror'' star. The field evolves in a fashion similar to a standing wave
(there are two main pulses repeatedly bouncing at the left and right
boundaries). It is only at the very late stages ($\eta \approx \pi$) that
amplification can be clearly seen, in agreement with the previous energy
considerations.

\section{A physical insight}

In this Section we give some physical arguments which will clarify the
apparent inefficiency of parametric amplification. Let us discuss first what
happens in the case of cosmological perturbations. There, we have two parameters
which regulate the strength of parametric amplification: the wavelength $\lambda$
of the perturbation and the Hubble radius $R_{\rm H} \equiv a/\dot{a} $. It is
well known \cite{lpg} that amplification is more efficient for large, ``superhubble'' modes,
i.e. $\lambda > R_{\rm H} $. This is easily understood, as it is a general property of wave 
propagation in a gravitational field that long wavelengths are the ones that are
dominantly affected (an example is provided by Fig ~\ref{t_ev2}).

\begin{figure}[tbh]
\centerline{\epsfysize=8cm \epsfbox{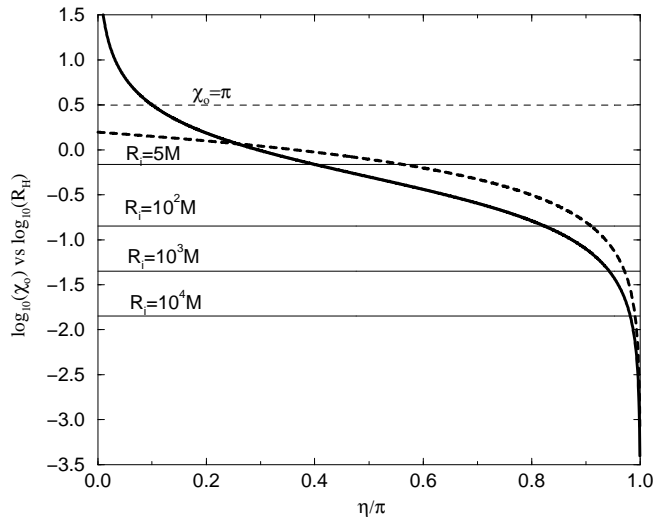}}
\caption{The amplification ``window''. The thick solid line corresponds
to one half the ``Hubble radius'' $R_{\rm H}$ as function of time, and 
each thin solid line represents the comoving radius of the star 
$\chi_{\rm o}(R_{\rm i})$. The thin
dashed line is plotted for $\chi_{\rm o}= \pi$ which corresponds to a pure
closed Friedmann spacetime. Finally, the thick dashed line is defined by 
eqn.~(\ref{cri2}) and represents the moment of time that the Schwarzschild 
horizon is crossed, for the different $R_{\rm i}/M$.} 
\label{hubble}
\end{figure}

In the case of a collapsing ``mirror'' star, there is a third lengthscale entering the 
problem, namely, the size $\chi_{\rm o}$ of the star. As evident from the solution
(\ref{solution2}), the field inside the star is made of the discrete spectrum
$\lambda_{\rm n}= 2\chi_{\rm o}/n$. Hence, the maximum wavelength is just equal
to the diameter of the star. For amplification to be significant, we should
demand that,
\begin{equation}
\chi_{\rm o} > \frac{1}{2} R_{\rm H}= \frac{1}{2\tan(\eta/2)}~.
\label{cri1}
\end{equation}
Moreover, there is another restriction set by the time required for the
star to cross its Schwarzschild radius. The condition $R \geq 2M $, with the help of
eqns. (\ref{Rad}),(\ref{RM}), gives,
\begin{equation}
\chi_{\rm o} <  \frac{\pi-\eta}{2}~.
\label{cri2}
\end{equation}
Conditions (\ref{cri1}) and (\ref{cri2}) are plotted together in
Fig.~\ref{hubble} for various values of $\chi_{\rm o}$, or equivalently, of  
$R_{\rm i}/M$. For a given choice of $R_{\rm i}/M$ there is an amplification ``window'' 
where there is at least one superhubble mode present. At this point we also need 
some input from the energy expression (\ref{energ2}), which says that the 
time-derivative of the potential is a crucial factor for amplification. 
Since $\partial_{\eta} V \sim \sin(\eta/2)/\cos^{3}(\eta/2) $ 
we could argue that amplification should become stronger as 
$\eta \to \pi$. In the same limit, and according to Fig.~\ref{hubble}, 
the amplification window shrinks considerably. In effect these two 
counter-balancing factors produce an almost uniform total amplification factor 
as it was shown in Fig.~\ref{figmir1}. On the other hand, for the case of 
pure Friedmann spacetime the corresponding window is significantly larger, 
which explains why amplification can be much stronger (as illustrated in Fig.~\ref{fig1}).


\section{Concluding discussion}

We have presented the results of an investigation on the possibility
of having parametrically amplified perturbations during gravitational
collapse. We have argued, based on the similarity between the time-varying 
gravitational field in the interior of a spherically collapsing star and the 
field of the closed Friedmann cosmological model, why this process will be present, 
at least in principle. Indeed, our investigation revealed that the process 
{\it is} present, and in order to obtain a quantitive estimate of the effect, 
we have constructed a  simplified collapse model (which captures the physics 
relevant to the problem) and studied time-evolutions of a scalar field 
placed inside the star. We have found that, unless the star has shrunk to a 
radius smaller than the Schwarzschild limit, the scalar field escapes from 
the star in a very short timescale for amplification to become important.
We have also quantified the amplification in terms of the field energy, and
set an upper limit (for scalar perturbations) by considering a perfectly 
reflecting star, i.e. a star that keeps the field trapped in its interior 
for the entire collapse event (this is what we called a ''mirror'' star). Unfortunately, we found that even under these very favourable conditions, 
amplification is only of order $\sim 1\% $ when the stellar surface is about 
to cross the star's Schwarzschild radius. It is only when the radius has shrunk 
to a much lower value that amplification becomes notable (for example, there is 
a $\sim 10\%$ amplification in the energy when $R=0.25M$). Moreover, these 
results remain almost invariant for a wide range of initial radii $R_{\rm i}/M$. 
We have given a simple physical argument, based on the comparison between the 
stellar radius, the Hubble radius $a/\dot{a}$ and the moment at which the
star crosses its Schwarzschild radius, to ``explain'' the low 
efficiency of the process.  

The results presented in this paper should encourage one to adopt a pessimistic
point of view regarding the astrophysical importance of parametric amplification
during gravitational collapse. However, this statement is far from being definite 
and must be viewed with some caution. We have considered one of the simplest 
possible models to describe a collapsing star. One could naturally ask how the 
above conclusion may change when more realistic models are adopted (which would incorporate features like pressure gradient, rotation etc.). In all these models, 
there is still present a time-varying background gravitational field which could 
pump energy into any kind of perturbations inside the star. There is no way at this 
point to say whether the amplification mechanism would operate with the same 
efficiency as in pressureless spherical collapse. Some steps towards including
rotational effects could be taken by considering a slowly rotating O-S collapse model
\cite{kegeles}. There are also other factors that should be taken into account 
in realistic collapse. For example, it is well established \cite{singh} that an inhomogeneous collapsing star can easily give birth to a naked singularity instead 
of a black hole. Under such conditions, our results suggest that strongly amplified fields could be generated while the stellar radius is $ < M $, and subsequently escape 
to infinity. In our view, and despite the ``negative'' results of this study, 
further (and more detailed) work is needed to give a more complete answer on the
astrophysical significance of parametric amplification.


\acknowledgements

We are grateful to L.P. Grishchuk for providing the idea that initiated 
this work and for many helpful discussions. We would also like to thank 
N. Andersson, C. Gundlach, J. Inglesfield and B.S. Sathyaprakash for useful 
discussions/comments.


\appendix

\section{Analytic solution for the ``mirror'' star}

In this Appendix we derive the analytic solution used to describe
a scalar field inside the ``mirror'' star of Section IV.
We rewrite eqn.~(\ref{master2}),
\begin{eqnarray}
\partial_{\eta}^2 \psi - \partial_{\chi}^2 \psi + V(\eta,\chi) \psi =0
\nonumber
\end{eqnarray}
and perform separation of variables:
\begin{equation}
\psi(\eta, \chi) = H(\eta)X(\chi).
\label{sep}
\end{equation}
We then get the following pair of equations for $H$ and $X$,
\begin{eqnarray}
\partial^2_{\eta}H +\left[ \nu^2 - \frac{1}{2}\cos^{-2}(\eta/2) \right]H &=& 0~,
\label{teq} \\
\partial^2_{\chi}X + \left[ \nu^2 - \frac{\ell(\ell+1)}{\sin^2(\chi)}\right]X &=& 0~.
\label{seq}
\end{eqnarray}
where $\nu$ is the separation constant. The solution of these equations
is,
\begin{eqnarray}
X_\nu(\chi)=c(\nu,l)\sqrt{\sin(\chi)} P^{-l-\frac1{2}}_{\nu-\frac1{2}}
[\cos(\chi)],
\label{Xsol} \\
\nonumber \\
H_\nu(\eta)=c_1(\nu)u_1(\nu,\eta)+c_2(\nu)u_2(\nu,\eta),
\label{Hsol} 
\end{eqnarray}
where $c, c_1, c_2$ are constants (with respect to $\eta$ and $\chi$),
$ P^{\mu}_{\kappa}[z]$ are associated Legendre functions and
\begin{eqnarray}
u_1=\cos[\nu(\pi-\eta)] -\frac1{2\nu}\tan(\eta/2)
\sin[\nu(\pi-\eta)], 
\\
u_2= 2\nu \sin[\nu(\pi-\eta)] + \tan(\eta/2)
\cos[\nu(\pi-\eta)].  
\end{eqnarray}
The above solutions already incorporate the ``left'' boundary condition
$\psi(\eta,0)=0$.

At this point, we restrict our attention to the monopole case ($\ell=0$) only.
The time dependence of the field $\psi$ (as given by eqn. (\ref{teq}) is the same 
irrespective of $\ell$, so consideration of $\ell \ne 0$ will just overcomplicate 
the analysis without providing any new physical information.

We next impose the ``right''  boundary condition, $\psi(\eta, \chi_{\rm o})=0$. 
For $\ell=0$ we find the following eigenvalues and  eigenfunctions:
\begin{eqnarray}
\nu_n=\frac{\pi}{\chi_0}n,~~ \mbox{ for }\;\;\; n=\pm 1,\pm 2, \dots  
\nonumber \\
\psi_n=\left[ c_1(\nu_n)u_1(\nu_n,\eta)+c_2(\nu_n)u_2(\nu_n,\eta) \right]
\sin\left(\frac{\pi n}{\chi_0}\chi\right)~.
\nonumber
\end{eqnarray}
The full field $\psi$ can be presented as an infinite sum of these 
eigenfunctions. The two remaining constants $c_1,c_2$ can be fixed
by the use of initial conditions (assumed static here). The 
final expression for $\psi$ is,
\begin{eqnarray}
\psi(\eta, \chi)=\frac{2}{\chi_0} \sum_{n=1}^{\infty} C_n
\left[\cos\left( \frac{\pi n}{\chi_0}\eta\right) +\frac{\chi_0}{2\pi n}
\tan\left(\frac{\eta}{2}\right)\sin\left( \frac{\pi n}{\chi_0}\eta\right)
\right] \sin\left(\frac{\pi n}{\chi_0}\chi\right)~,
\label{mirsol}
\end{eqnarray}
where
\begin{equation}
C_n=\int^{\chi_0}_{0} d\chi~ \psi_0(\chi)
\sin\left(\frac{\pi n}{\chi_0}\chi\right)~.
\end{equation}
This solution exhibits a standing wave-like behaviour. Moreover,
by setting $\chi_{\rm o}=\pi$ it can be used to describe 
scalar wave propagation in a closed Friedmann spacetime.
When applying  (\ref{mirsol}) in practice, we found that the
$n$-sum is rapidly convergent (typically, no more than 20-25 terms are required).


\section{Energy ``conservation'' law for the Klein-Gordon equation}

The ``master'' equation that governs the field $\psi$ is 
a Klein-Gordon equation. Here we intend to derive a simple 
``energy conservation'' expression for this type of equations
when homogeneous boundary conditions are assumed for the field
at the right and left endpoints. 

Consider again eqn.~(\ref{master2})
\begin{equation}
\partial^2_{\eta}\psi-\partial^2_{\chi}\psi + V(\eta,\chi)\psi=0,
\label{aeq1}
\end{equation}
where $V(\eta,\chi)$ is an effective potential which can
generally have both spatial and time dependence. 
Suppose that the field is confined in the region 
$0\leq \chi \leq \chi_{\rm o}$ and $\psi(\eta,0)=\psi(\eta,\chi_{\rm o})=0$ 
(as in the model of the ``mirror'' star of Section IV). 
Multiplying eqn.~(\ref{aeq1}) with $2\partial_{\eta}\psi$ and making 
a trivial rearrangement we obtain
\begin{eqnarray}
2\partial_{\eta}\psi\partial^2_{\eta}\psi - 
2\partial_{\chi}(\partial_{\eta}\psi\partial_{\chi}\psi) +
2\partial_{\chi}\partial_{\eta}\psi\partial_{\chi}\psi + 
2V(\eta,\chi)\psi\partial_{\eta}\psi=0.
\nonumber
\end{eqnarray}
This can also be written as
\begin{equation}
\partial_{\eta} \left[ (\partial_{\eta}\psi)^2 + (\partial_{\chi}\psi)^2 +
V\psi^2 \right] - \psi^2\partial_{\eta}V - 2\partial_{\chi}
(\partial_{\eta}\psi\partial_{\chi}\psi)=0.
\label{aeq2}
\end{equation}
Integrating eqn.~(\ref{aeq2}) from $0$ to $\chi_0$  and taking into account
the boundary conditions we get
\begin{equation}
\partial_{\eta}\int_{0}^{\chi_0} \left[ (\partial_{\eta}\psi)^2 + 
(\partial_{\chi}\psi)^2 + V\psi^2 \right] d\chi =
\int_{0}^{\chi_0} d\chi\psi^2\partial_{\eta}V.
\label{aeq3}
\end{equation}
We define the quantity
\begin{equation}
E(\eta)\equiv \frac 1{2} \int_{0}^{\chi_0} \left[ (\partial_{\eta}\psi)^2 + 
(\partial_{\chi}\psi)^2 + V\psi^2 \right] d\chi.
\end{equation}
as the field's total energy (this can be compared with the energy
expression for a vibrating string). We have then shown that,
\begin{equation}
E(\eta) -E(0)= \frac{1}{2} \int_{0}^{\eta} \int_{0}^{\chi_{\rm o}} 
d\eta^{\prime} d\chi~ \psi^2 \partial_{\eta^{\prime}} V.
\end{equation}
In the ``usual'' case of time-independent potentials, this equation
just says that the field energy is conserved. This property is spoiled
when the potential is allowed to vary in time due to the
coupling of the field and the time-derivative of the potential.   
This simple relation is an alternative way to view amplification
of fields by time-dependent potentials.


\end{document}